\documentclass{aa}
\usepackage{graphics}
\usepackage{latexsym}

\begin{document}


   \title{H{\sc i} deficiency in the galaxy cluster ACO~3627}

   \subtitle{ATCA\thanks{The Australia Telescope Compact Array (ATCA)
is operated by the Australia Telescope National Facility, CSIRO, as a National
 Research Facility.} observations in the Great Attractor region}

   \author{B.~Vollmer\inst{1,2,8}, V.~Cayatte\inst{1}, W.~van Driel\inst{1,3}, P.A.~Henning\inst{4}, R.C.~Kraan-Korteweg\inst{1,5}, C.~Balkowski\inst{1}, P.A.~Woudt\inst{6,7} \and W.J.~Duschl\inst{2,8}}

   \offprints{B.~Vollmer, e-mail: bvollmer@mpifr-bonn.mpg.de}

   \institute{Observatoire de Paris, DAEC,
              UMR 8631, CNRS et Universit\'e Paris 7,
	      F-92195 Meudon Cedex, France. \and
	      Institut f\"ur Theoretische
              Astrophysik der Universit\"at Heidelberg,
	      Tiergartenstra{\ss}e 15, D-69121 Heidelberg, Germany. \and
	      Unit\'e Scientifique Nan\c{c}ay, CNRS USR 704B, Observatoire de
	      Paris, F-92195 Meudon Cedex, France. \and
	      Institute for Astrophysics, University of New Mexico,
	      800 Yale Boulevard, NE, Albuquerque, NM 87131, USA. \and
	      Departamento de Astronom{\'{\i}}a, Universidad de Guanajuato,
	      Apdo. Postal 144, Guanajuato, Gto. 36000, Mexico. \and
	      Department of Astronomy, University of Cape Town,
	      Rondebosch, 7700 South Africa. \and
              European Southern Observatory, Karl-Schwarzschild-Str. 2,
	      D-85748 Garching, Germany. \and
	      Max-Planck-Institut f\"ur Radioastronomie, Auf dem H\"ugel 69,
	      D-53121 Bonn, Germany.
	      }

   \date{Received / Accepted}

   \authorrunning{Vollmer et al.}

\abstract{ATCA 21 cm H{\sc i} observations of the rich galaxy cluster ACO~3627
in the Great Attractor region are presented. Three fields of 30$'$
diameter located within one Abell radius of ACO~3627 were
observed with a resolution of 15$''$ and an rms noise of $\sim$1
mJy/beam. Only two galaxies were detected in these fields. We compare
their H{\sc i} distribution to new optical R-band images and discuss
their velocity fields. The first galaxy is a gas-rich unperturbed
spiral whereas the second shows a peculiar H{\sc i} distribution.
The estimated 3$\sigma$ H{\sc i} mass limit of our observations is
$\sim 7\,10^{8}$ M$_{\odot}$ for a line width of 150 km\,s$^{-1}$.
The non-detection of a considerable number of luminous spiral galaxies
indicates that the spiral galaxies are H{\sc i} deficient. The low
detection rate is comparable to the H{\sc i} deficient Coma cluster
(Bravo-Alfaro et al. 2000). ACO~3627 is a bright X-ray cluster. We
therefore suspect that ram pressure stripping is responsible for
the H{\sc i} deficiency of the bright cluster spirals.
\keywords{
Galaxies: interactions -- Galaxies: ISM -- Galaxies: kinematics and dynamics}
}


\maketitle

\section{Introduction}
Spiral galaxies in clusters are H{\sc i} deficient compared to similar
galaxies (equivalent size and morphological type) in the field (see,
e.g., Chamaraux et al. 1980). In the Virgo cluster, the H{\sc i} gas is
stripped in the outer regions of spiral galaxies, an effect which is
more pronounced for central spirals (Cayatte et al. 1990).  The same
tendency was observed for galaxies in the Coma cluster (Bravo-Alfaro et
al. 2000).
H{\sc i} deficiency seems correlated with the X-ray luminosity of clusters
(Giovanelli \& Haynes 1985; see however Solanes et al. 2000).

In this context, the cluster ACO~3627 provides an excellent probe for
further studies of environmental effects. ACO~3627 has recently been
identified as a nearby massive galaxy cluster behind the Milky Way
(Kraan-Korteweg et al. 1996). Its position in velocity space of
($l,b,<v>) = (325.3^{\rm o}, -7.2^{\rm o}, 4844~$km\,s$^{-1}$) --
only $\sim$9$^{\rm o}$ from the predicted center of the Great Attractor
(Kolatt et al. 1995)
-- makes ACO~3627 the most massive known cluster in the Great Attractor
region. It is comparable in cluster mass and richness to the Coma
cluster (Woudt 1998) but located one and a half times closer to us.

ACO~3627 also has an X-ray luminosity comparable to that of the Coma
cluster (B\"{o}hringer et al. 1996). Its X-ray distribution is not
spherically symmetric, but shows indications of an ongoing cluster
merger (B\"{o}hringer et al. 1996, Tamura et al. 1998). We therefore
expect that its spiral galaxies might be H{\sc i} deficient -- similar
to the spirals of the Coma cluster. This suspicion and the existence
of a fairly large number of spiral galaxies make this cluster an ideal
candidate for H{\sc i} deficiency studies of its cluster population.

The proximity of ACO~3627 to the Galactic plane impedes a complete and
unbiased study of ACO~3627 in the optical (Woudt 1998). 21~cm
observations are, however, not affected by the dust in the cluster
area.  Various pointed H{\sc i} observations have been obtained for the
optically identified spiral galaxies in the ACO~3627 area to noise
levels of typically 3-5~mJy (Kraan-Korteweg et al. 1997).
In addition, the southern Zone of Avoidance is being searched
systematically for galaxies in H{\sc i} with the multi-beam receiver
on the 64~m Parkes radiotelescope (cf. Henning et al. 1999,
Henning et al. 2000 for results from the
shallow survey, and Kraan-Korteweg \& Juraszek 2000 for first results
in the Great Attractor region from the full-sensitivity survey).  As
this survey is limited to Galactic latitudes of $|b|<5.25^{\rm o}$ it
covers, however, only a very small fraction of the ACO~3627 cluster
within its Abell radius.

To complement the above described single dish observations, we
obtained detailed 21~cm line images using the Australia  Telescope
Compact Array (ATCA). Three fields were observed within the Abell radius of
the cluster ACO~3627, plus an additional field in the outskirts of the
cluster (to a noise level of about 1 mJy/beam) to study the effects of
the cluster environment on its members in different parts of the
cluster.  Unfortunately, one of the strongest known extragalactic
radio sources is located at the center of the cluster, i.e. the
wide-angle-tail (WAT) radio galaxy B1610--608 (see Jones \& McAdam
1996 for a detailed discussion), making the analysis of the data of
the central ATCA field quite difficult.

The observations and the data reduction procedures are presented in
Sect.~2.  The H{\sc i} detections are shown and discussed in Sect.~3.
In Sect.~4, we compare our detection rate with the detection rate
of the Coma cluster as derived from VLA observations (Bravo-Alfaro et
al. 2000). The conclusions are given in Sect.~5.

We assume a distance of 79\,h$_{50}^{-1}$ Mpc for all galaxies in the
ACO~3627 cluster as derived from the Tully-Fisher relation by Woudt
(1998).

\section{Observations and data reduction}

\begin{figure}
	\resizebox{\hsize}{!}{\includegraphics{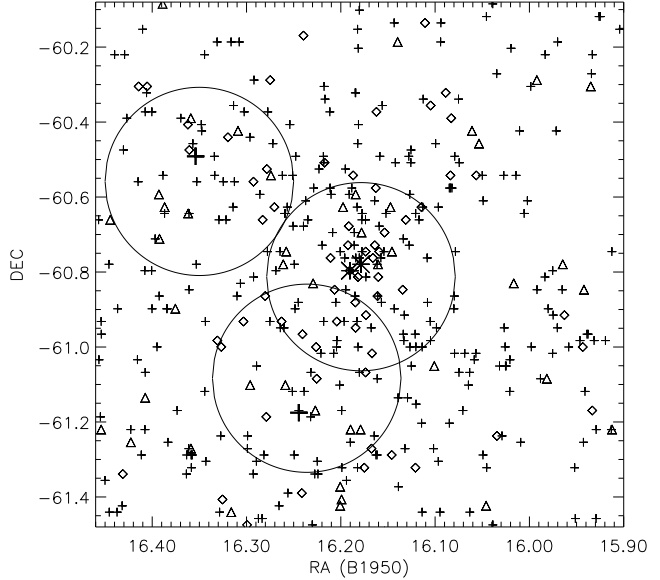}}
	\caption{Overview of the three inner observed ATCA fields.
	The central cD galaxies are marked as stars in the middle of the map.
	Crosses indicate spiral galaxies, diamonds ellipticals or lenticulars,
	and triangles unclassified galaxies. The observed fields
	are outlined by circles. The locations of the two detected spiral
	galaxies are marked as large, bold crosses.}
	\label{fig:allgal}
\end{figure}

We observed three
fields inside the Abell radius of ACO~3627 (cf. Fig.~\ref{fig:allgal}) and one
comparison field well outside the cluster, each in 4 different
ATCA configurations (Table~\ref{tab:obstime}).

\begin{table}
     \caption{Observations -- ATCA Configurations}
         \label{tab:obstime}
      \[
         \begin{array}{lll}
            \hline
            \noalign{\smallskip}
            {\rm Configuration}\ \  & {\rm begin} & {\rm end} \\
            \noalign{\smallskip}
            \hline
            0.375 & 1996\ {\rm September}\ 25 & 1996\ {\rm September}\ 30\\
            0.75{\rm A} & 1996\ {\rm November}\ 16 & 1996\ {\rm November}\ 19\\
            1.5{\rm A} & 1996\ {\rm October}\ 18 & 1996\ {\rm October}\ 22\\
	    6.0{\rm B} & 1996\ {\rm September}\ 4 & 1996\ {\rm September}\ 7\\
            \noalign{\smallskip}
            \hline
         \end{array}
      \]
\end{table}
The observing time for each field was $\sim$12 hr per configuration.
The baselines employed spanned the range from 31 m to 6000 m.
Details of the observations can be found in Table~\ref{tab:detobs}.
The first column is the name of the field followed by the coordinates
of the field center, the HPBW ($\alpha \times \delta$) of the synthesized
beam, and the rms noise.
\begin{table}
     \caption{Observations --Fields}
         \label{tab:detobs}
      \[
         \begin{array}{lcccc}
            \hline
            \noalign{\smallskip}
            {\rm field} & {\rm RA(2000)} & {\rm DEC(2000)} & {\rm HPBW} & {\rm rms\ noise}\\
	      & & & & {\rm ({\rm mJy/beam})}\\
            \noalign{\smallskip}
            \hline
            \noalign{\smallskip}
            {\rm Comp.} & 16$$^{\rm h}$$51$$^{\rm m}$$17$$^{\rm s}$$5 & -58$$^{\rm o}$$31$$'$$31$$''$$ & $$10''\times 8''$$ & 1.0 \\
	    {\rm East} & 16$$^{\rm h}$$25$$^{\rm m}$$22$$^{\rm s}$$0 & -60$$^{\rm o}$$40$$'$$21$$''$$ & $$15''\times 14''$$ & 1.0 \\
	    {\rm South-East} & 16$$^{\rm h}$$18$$^{\rm m}$$32$$^{\rm s}$$9 & -61$$^{\rm o}$$12$$'$$18$$''$$ & $$14''\times 12''$$ & 1.2 \\
	    {\rm Center} & 16$$^{\rm h}$$15$$^{\rm m}$$10$$^{\rm s}$$9 & -60$$^{\rm o}$$56$$'$$16$$''$$ & $$15''\times 10''$$ & 1.3 \\
            \noalign{\smallskip}
            \hline
         \end{array}
      \]
\end{table}
The effective size of each field is the individual antenna HPBW=30$'$.
We used a
bandwidth of 32 MHz (6900 km\,s$^{-1}$), divided into 256 channels of
125 kHz (27 km\,s$^{-1}$) each and centered on a frequency of 1.397 GHz
(corresponding to an heliocentric velocity of 5023 km\,s$^{-1}$).
Thus, the effective velocity coverage ranges from 1530 km\,s$^{-1}$
to 8510 km\,s$^{-1}$ with a spectral resolution of $\sim 30$ km\,s$^{-1}$.
The primary ATCA calibrator, PKS B1934--638, was observed at least once
per 12 h run. Each 40 minute observation of a field was bracketed by short
observation of a nearby secondary calibrator: PKS B1718--649.
Each $uv$ database for a given configuration and a given field
was processed separately using the NRAO Astronomical Image Processing
System (AIPS) package. The subtraction of the continuum was found to be
the main difficulty for the cluster fields due to the presence of the very
strong [55 Jy] continuum source, B1610-608. This will therefore be discussed
in detail.

(i) Comparison field: As there is a point source with a flux density of
$\sim$180 mJy near the field center, we CLEANed  and subtracted the CLEAN
components
directly in the $uv$ plane (AIPS task: UVSUB). The remaining continuum
was then removed by linear $uv$ interpolation with respect to
frequency (AIPS task: UVLIN).

(ii) Eastern field: We made a CLEANed image of the WAT galaxy, B1610--608
shifting the phase center to coincide with the center of this galaxy
and subtracted the CLEAN components directly in the $uv$ plane.
In a second step another three fainter point sources were subtracted in
the way described in (i). The remaining continuum was then removed by a
linear $uv$ interpolation with respect to frequency.

(iii) South-eastern field: As B1610--608 lies just beyond the upper western
edge of the image, its sidelobes dominate the image. We reconstructed
this very extended ($\sim$14$'$) source shifting the phase center
again to coincide with the galaxy center and subtracting the CLEAN
components directly in the $uv$ plane.
Two point sources were then removed using the technique
described in (i). As the sidelobes of B1610--608 then still dominated the
image we repeated the procedure removing the WAT galaxy. At the end
two linear $uv$ interpolation procedure were applied, the first with a
shift on B1610--608, the second with a zero shift.

(iv) Central field: B1610--608 is located in the center of this field.
We reconstructed this source and subtracted the CLEAN components
directly in the $uv$ plane. The CLEANed flux is 57 Jy
for configuration 0.375 and still 55 Jy for configuration 0.75A.
The main problem in this field is that the frequency baselines
are not flat and even show some discontinuities.
The subtraction of a polynomial fit of order 4 in the image cube
(AIPS task: IMLIN)
resulted in an acceptable rms noise which is only a somewhat higher
($\sim 30\%$) than those attained in the other fields.

After combining the databases of the four configurations for each field,
images were obtained using the AIPS task IMAGR applying different
weightings. In the case of uniform weighting this resulted in a final
resolution of 15$''$ and rms noises
of $\sim$1 mJy/beam. As expected, the rms noise in the different
fields increases with decreasing distance to B1610--608.

\section{Results}

In order to detect possible galaxy candidates, we applied two
different strategies.

Since the optical galaxy diameters are known to range from 20$''$ to 100$''$,
we analysed our data at different resolutions. We used the full
resolution data as a search for an optimized study of small size galaxies.
For an optimized study of larger objects ($\geq 30''$) we degraded
the resolution to about 30$''$, applying a $uv$ taper of 10k$\lambda$ and
a weighting close to a natural weighting, resulting in beamsizes of about
$30''\times30''$. This increases the sensitivity to extended sources with
diameters greater than 15$'' \simeq 5$ kpc by factors up to 4 compared with
uniform weighting.
The main galaxy search was made in a data cube with a degraded
resolution of $\sim 30''$.
First, for each of the four fields we added the emission of all velocity
channels at this resolution above a threshold of 3$\sigma$.
We then made spectra for all peaks in the integrated H{\sc i} maps
exceeding the 3$\sigma$ level.

In order to double check the results of these two search techniques
we made small sized data cubes ($2'\times2'\times250$ km\,s$^{-1}$)
centered on the position and velocity of all Woudt (1998) galaxies
in the fields. The relatively small size of the cubes allow a simple,
reliable continuum subtraction. We found no additional detections
using this method.

Woudt (1998) identified $\sim$80 spiral galaxies in the three fields,
out of which 20 have extinction-corrected blue luminosities
B$_{\rm J}^{0} < 15$ mag.
In total, we found two galaxies, one in eastern field and one
in south-eastern field.
Their properties are listed in Table~\ref{tab:galaxies}.
The columns are: (1) galaxy name (2) RA(2000) of the
optical center (3) DEC(2000) of the optical center (4) diameter of
the major axis (5) diameter of the minor axis (6) extinction-corrected
diameter of the major axis (7) extinction-corrected apparent B magnitude
(the extinctions were taken from the Schlegel et al. 1998 maps)
(8) range of morphological type
(9) optical heliocentric velocity (10) central H{\sc i} velocity
(11) H{\sc i} linewidth at 20\%
(12) H{\sc i} mass (13) M$_{\rm HI}$/L$_{\rm B}$.
\begin{table*}
   \caption{Detected spiral galaxies.}
         \label{tab:galaxies}
      \[
         \begin{array}{lcccccccccccc}
	  \hline
          \noalign{\smallskip}
          {\rm Name} & {\rm RA(2000)} & {\rm DEC(2000)} & {\rm D} & {\rm d} & {\rm D_{0}} & {\rm B_{\rm J}^{0}} & {\rm type}
 & {\rm v}_{\rm opt} & {\rm v}_{\rm HI} & {\rm W}_{20} & {\rm M_{\rm HI}} & {\rm M_{\rm HI}/L_{\rm B}}\\
	    &     &     &($$''$$)&($$''$$)&($$''$$)&({\rm mag})& &({\rm km\,s^{-1}})&({\rm km\,s^{-1}}) & ({\rm km\,s^{-1}}) &(10^{9}\ {\rm M_{\odot}})& ({\rm M}_{\odot}/{\rm L}_{\odot, \rm B})\\	
	(1) & (2) & (3) & (4) & (5) & (6) & (7) & (8) & (9) & (10) & (11) & (12)& (13) \\
          \noalign{\smallskip}
	  \hline
          \noalign{\smallskip}
	  {\rm WKK}6801 & 16\ 25\ 36.2 & -60\ 36\ 19 & 38 & 16 & 47 & 15.7 & {\rm Sdm-Im} & 3518$$\pm$$7 & 3529$$\pm$$30 & 160 & 13.1 & 2.43 \\
	  {\rm WKK}6489 & 16\ 19\ 03.7 & -61\ 17\ 47 & 48 & 13 & 61 & 15.1 & {\rm Sb-Sd} & 3794$$\pm$$58 & 3880$$\pm$$30 & 140 & 1.6 & 0.17 \\
          \noalign{\smallskip}
          \hline
         \end{array}
      \]
\end{table*}
For further analysis we made images of the best possible spatial resolution
(15$''$) for the detected galaxies. For the determination of integrated
spectra and the total flux we use the 30$''$ resolution cubes which are
more sensitive for these extended sources.
We will now discuss the detected galaxies in detail.

\subsection{WKK~6801}

\begin{figure}
	\resizebox{\hsize}{!}{\includegraphics{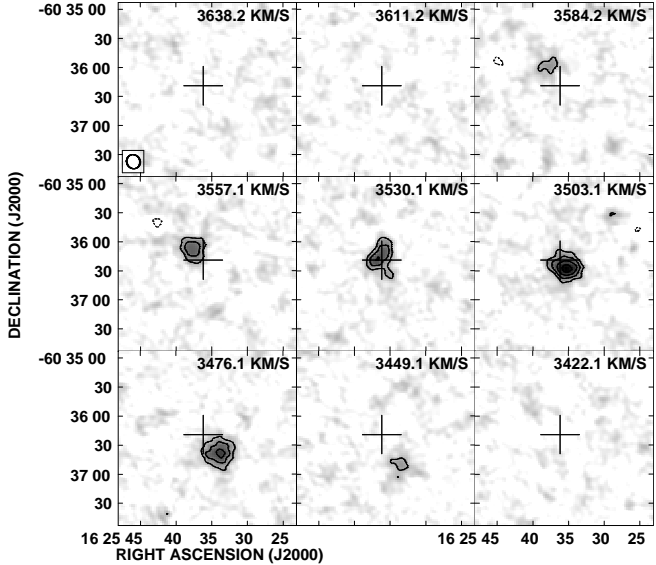}}
	\caption{Continuum subtracted channel maps of WKK~6801.
	The contours are -2.79, 2.79 (3$\sigma$), 4.65, 6.50, 8.36 mJy/beam.
	The resolution is 15$''$. The cross marks the optical center.}
	\label{fig:channels6801}
\end{figure}

\begin{figure}
	\resizebox{\hsize}{!}{\includegraphics{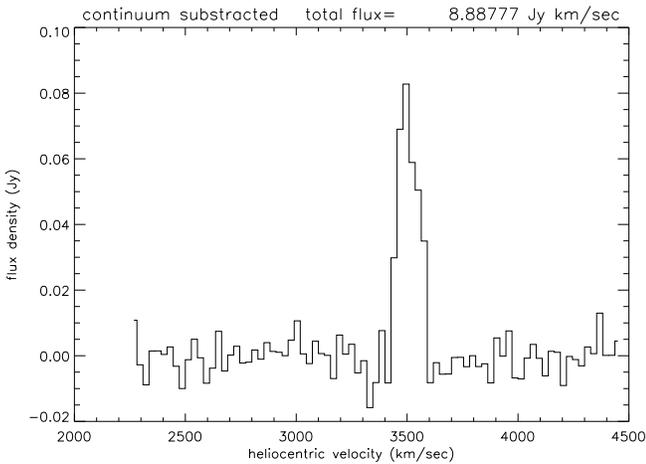}}
	\caption{The integrated spectrum (over a 60$'' \times 60''$ area)
	of WKK~6801. The channel width is 27 km\,s$^{-1}$.
	The beam size is 30$''$.}
	\label{fig:spect6801}
\end{figure}

\begin{figure*}
	\resizebox{\hsize}{!}{\includegraphics{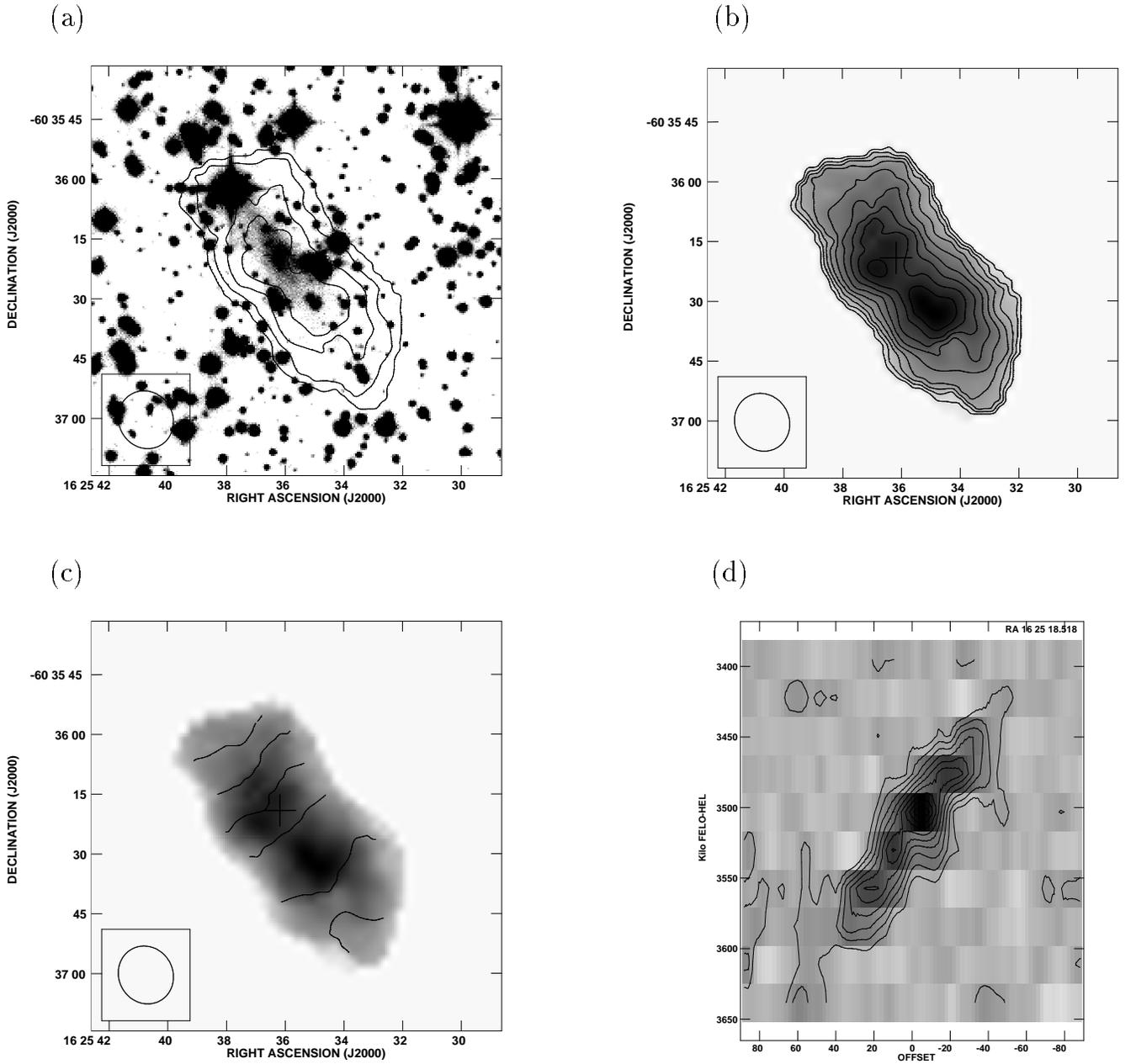}}
	\caption{Galaxy WKK~6801.
	(a) Overlay of a high contrast optical R-band
	image and the H{\sc i} distribution with some selected contours.
	The H{\sc i} beam is shown in the lower left corner.
	(b) H{\sc i} distribution. The outer contour
	corresponds to 2.3\,10$^{20}$ cm$^{-2}$, the
	contour steps have the same value. The 15$''$ beam is shown in
	the bottom left corner. The cross marks the optical center.
	(c) Velocity field. The first contour in the north-east
	corresponds to 3560 km\,s$^{-1}$. The contours are in steps
	of -20 km\,s$^{-1}$. The cross marks the optical center.
	(d) Position-velocity plot along the major axis.
	Negative position offsets correspond to the
	south-west direction. Contour levels are (1, 2, 3, 4, 5, 6, 7, 8, 9)
	$\times$ 0.925 mJy/beam.}
	\label{fig:wkk6801_all}
\end{figure*}

This H{\sc i} detection in the field East coincides with the optical
identification WKK~6801 (see Fig.~\ref{fig:wkk6801_all}a).
It is located at a projected distance of 2.6$^{\rm o}$=3.5 Mpc from
the cluster center.
This spiral galaxy has an extinction-corrected B magnitude of
${\rm B}_{\rm J}^{0}$=15.7, an extinction-corrected diameter of
${\rm D}_{0}=47''$, and a radial velocity of 3519$\pm$7 km\,s$^{-1}$
(Woudt 1998).
Fig.~\ref{fig:channels6801} shows the full resolution (15$''$) channel maps.
The galaxy is visible over a range of 6 channels.
The integrated H{\sc i} spectrum measured over a 60$'' \times 60''$
area centered on the optical center, using the 30$''$ resolution data can be
seen in Fig.~\ref{fig:spect6801}. It does not show the usual double horn
feature of inclined disk galaxies but only a single peak.
The derived total line flux corrected for beam attenuation is
8.9 Jy\,km\,s$^{-1}$.
This corresponds to a total H{\sc i} mass of 1.3\,10$^{10}$ M$_{\odot}$
(for a distance of 79 Mpc).
The velocity channel separation is 27 km\,s$^{-1}$. We measured full widths
at, respectively, 20\% and 50\% of the maximum profile height of
$W_{20}$=160 km\,s$^{-1}$ and $W_{50}$=110 km\,s$^{-1}$.
The error on the width is of the order of the channel separation.
H{\sc i} spectra of extreme late type galaxies can show this kind of profile
(Matthews et al. 1998) with comparable linewidths.

The H{\sc i} distribution map can be seen in Fig.~\ref{fig:wkk6801_all}(b).
The galaxy has a symmetric H{\sc i} distribution in the outer parts of its
highly inclined disk,
whereas the inner part shows two asymmetric maxima with respect to the
optical galaxy center. The main maximum is located in the south-west.
The maximum surface density in the center
is 38 M$_{\odot}$/pc$^{2}$. The profile corresponds to those of
field spirals which do not show signs of perturbation in their
H{\sc i} content (Cayatte et al. 1994). The derived H{\sc i} diameter
at a column density of 2.3\,10$^{20}$ cm$^{-2}$ is
$D_{HI}=82''$=31 kpc, leading to a mean H{\sc i} surface density of
17 M$_{\odot}$/pc$^{2}$. Woudt (1998) gives an extinction-corrected optical
diameter of 47$''$ (E$_{({\rm B-V})}$=0.199; Schlegel et al. 1998).
This means that the H{\sc i} diameter exceeds the optical diameter by
almost a factor 2.

The velocity field is shown in Fig.~\ref{fig:wkk6801_all}(c).
Its regular form is only perturbed at the location of the south-western
maximum which could be a spiral arm.
The position-velocity plot along the major axis can be seen in
Fig.~\ref{fig:wkk6801_all}(d). It shows a linearly rising rotation
curve which is perturbed in the galaxy centre.
We fitted a rotation curve to the data excluding $\pm 30^{\rm o}$ around
the minor axis. The systemic velocity is 3518$\pm$7 km\,s$^{-1}$.
The kinematical center coincides within 2$''$ with the optical
center given by Woudt (1998). The position angle is PA=35$^{\rm o}$,
the inclination i$\simeq$60$^{\rm o}$, and the maximum rotation velocity
$v_{\rm max} \simeq$75 km\,s$^{-1}$.

The galaxy has an extinction-corrected magnitude of
$B_{\rm J}^{0} \simeq$ 15.7 (Woudt 1998).
This gives an absolute magnitude of $M_{B} \simeq -18.8$ leading
to an H{\sc i} mass-to-light ratio $M_{HI}/L_{B} \simeq$ 2.43
M$_{\odot}$/L$_{\odot, \rm B}$ (using
a distance of 79 Mpc). This corresponds to the median value for an Sd or Sdm
galaxy (Roberts \& Haynes 1994). The maximum velocity of the rotation curve
of 75 km\,s$^{-1}$
and the line profile are also representative of a very late type galaxy.

\subsection{WKK~6489}

\begin{figure}[htb]
\resizebox{\hsize}{!}{\includegraphics{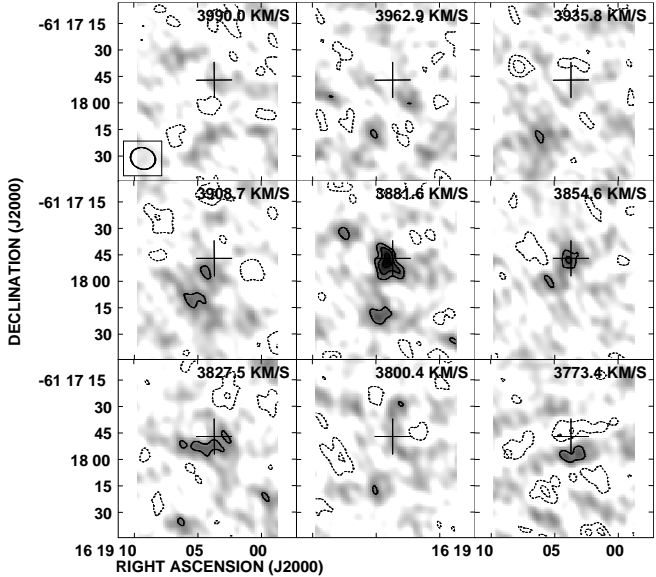}}
\caption{The channel maps of WKK~6489. The contours are
-2.39, -1.20, 2.39 (3$\sigma$), 3.34, 4.30 mJy/beam.
The cross marks the optical center. The resolution is 15$''$.}
\label{fig:channels6489}
\end{figure}

\begin{figure}[htb]
\resizebox{\hsize}{!}{\includegraphics{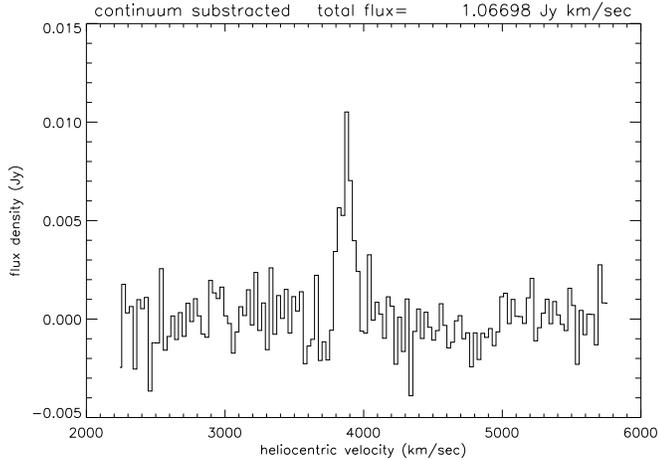}}
\caption{The integrated spectrum of WKK~6489
over a 45$'' \times 45''$ area.
The channel width is 27 km\,s$^{-1}$. The beam size is 30$''$.}
\label{fig:spect6489}
\end{figure}

This H{\sc i} detection in the south-eastern field matches with the optical
identification of WKK~6489 (Fig.~\ref{fig:wkk6489_all}a).
This spiral galaxy has an extinction-corrected B magnitude of
${\rm B}_{\rm J}^{0}$=15.1, an
extinction-corrected diameter of ${\rm D}_{0}=61''$, and a velocity of
3794$\pm$58 km\,s$^{-1}$ (Woudt 1998).
A very weak detection is seen in co-added 2MASS near-infrared
(J,H,K-band) images, whose center lies only 2$''$ NW of Woudt's
optical position. Its estimated total magnitude in K is about 14.5-15,
no near-infrared colors could be derived.
Its apparent major axis diameter is about 20$''$, and its orientation and
axial ratio appear consistent with the R-band image.
The channel maps are shown in Fig.~\ref{fig:channels6489}. The galaxy
is clearly detected in 4 channels.
The H{\sc i} spectrum integrated over an area of $45'' \times 45''$
(Fig.~\ref{fig:spect6489}) shows that this galaxy has a
21 cm line flux which is 10 times smaller than that of WKK~6801.
The derived total flux is 1.07 Jy\,km\,s$^{-1}$, corresponding to a total
H{\sc i} mass of 1.6\,10$^{9}$ M$_{\odot}$.

WKK~6489 has a corrected magnitude of $B_{\rm J}^{0} \simeq$ 15.1
(Woudt 1998).
This corresponds to an absolute magnitude of $M_{B} \simeq -19.4$ mag
leading to an H{\sc i} mass-to-light ratio $M_{HI}/L_{B} \simeq$ 0.17
M$_{\odot}$/L$_{\odot, \rm B}$.

The channel width is 27 km\,s$^{-1}$, the beam size used here is 30$''$.
The measured full widths at, respectively, 20\% and 50\% of the
maximum profile height are $W_{20}=140$ km\,s$^{-1}$
and $W_{50}$=80 km\,s$^{-1}$.
Fig.~\ref{fig:wkk6489_all}(b) shows the H{\sc i} brightness distribution.
The cross indicates the optical center determined by Woudt (1998).
The main H{\sc i} emission lies south-east of the optical galaxy center.
Its velocity coincides with the optical velocity within the errors.
A second blob is located
at a distance of 30$''$ under the south-eastern end of the major axis.
Woudt (1998) gives
an extinction-corrected optical diameter of 61$''$. This means that
the south-eastern blob lies at the edge of the optical disk.
The velocity field can be seen in Fig.~\ref{fig:wkk6489_all}(c).
Its peculiar behaviour could partly be due to the high noise level
in the data.
The position-velocity along the major axis is shown in
Fig.~\ref{fig:wkk6489_all}(d).
We observe a continuous velocity field between the two distinct H{\sc i}
emission blobs.
Thus, the two blobs are possibly connected but our data does not permit
to draw a firm conlusion.
In principle the observed asymmetric H{\sc i} morphology could be due to:

\begin{figure*}[htb]
\resizebox{\hsize}{!}{\includegraphics{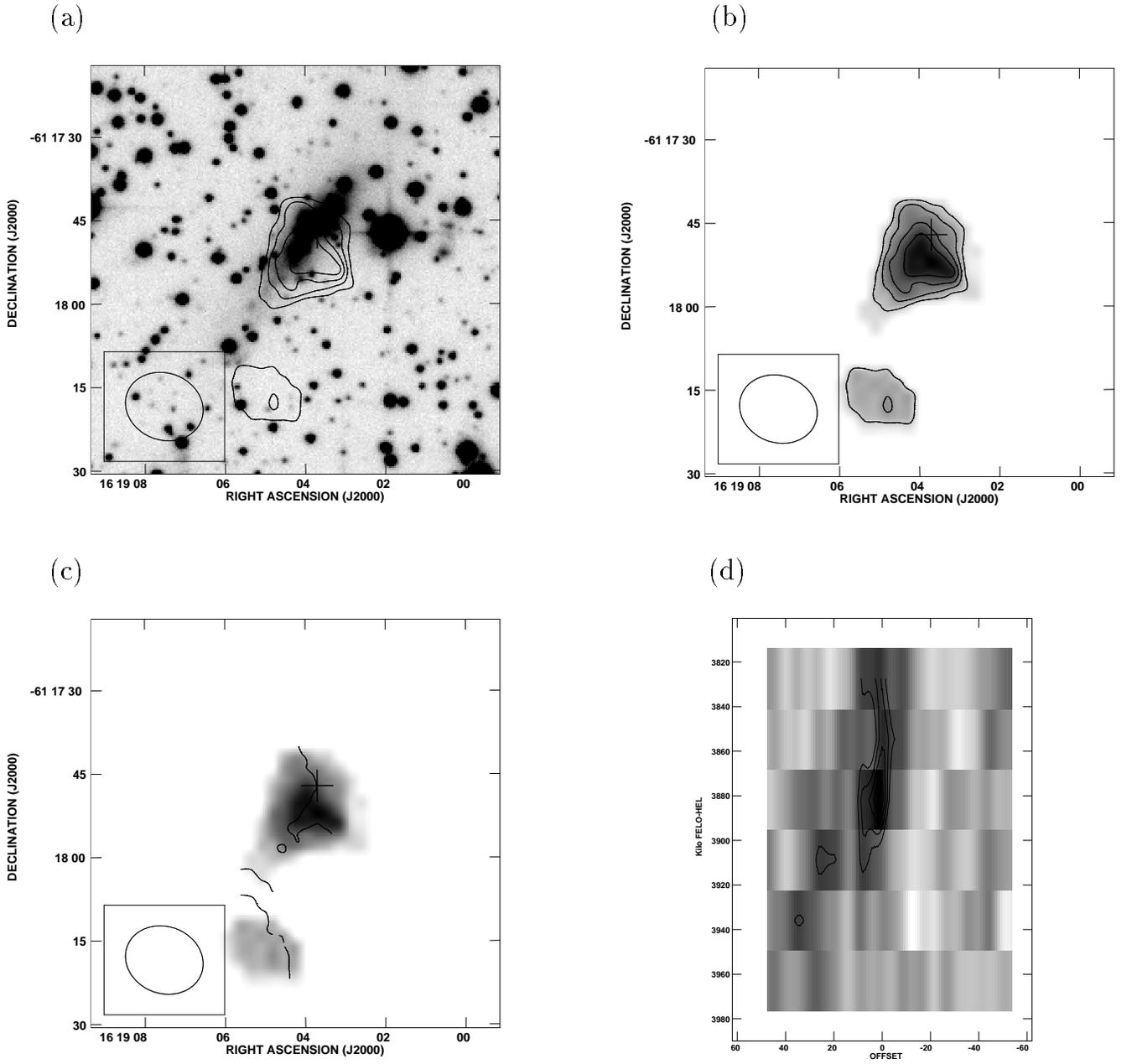}}
\caption{WKK~6489. (a) Overlay of a high contrast optical
R-band image and the H{\sc i} distribution.
(b) H{\sc i} distribution. The outer contour
corresponds to 1.7\,10$^{20}$ cm$^{-2}$,
the contour steps are 1.85\,10$^{21}$ cm$^{-2}$.
The beam is shown in the bottom left corner.
The cross marks the optical center.
(c) Velocity field. The first contour in the south-east corresponds
to 3880 km\,s$^{-1}$. The next contours are 3900,
3880, and 3860 km\,s$^{-1}$ northwards.
The cross marks the optical center.
The beam is shown in the upper right corner.
(d) Position-velocity plot along the major axis.
Negative position offsets correspond to the
north-west direction.}
\label{fig:wkk6489_all}
\end{figure*}

\begin{itemize}
\item
a galaxy--galaxy interaction,
\item
an interaction between the cluster's gravitational potential,
\item
ram pressure exerted by the hot intracluster medium on the fast moving
galaxy (Gunn \& Gott 1972).
\end{itemize}
Since there is no optical companion, a gravitational
interaction is unlikely. Moreover, its asymmetric H{\sc i} gas
distribution reminds that of NGC~4848 located in Coma
(Bravo-Alfaro et al. 2000), a cluster which is very similar to ACO~3627
(see Sect.~\ref{sec:coma}).
N-body models simulating the gas disk of spiral galaxies falling into
the Virgo cluster (Vollmer et al. 2001) show the importance of ram pressure
stripping in this cluster. We expect this effect to be still more
pronounced in bigger clusters as ACO~3627 or Coma.
We therefore think that ram pressure is the
most probable mechanism which caused the observed asymmetry.

\subsection{Number of detections}

The 3$\sigma$ detection limit in each field is $\sim$3 mJy/beam
in one velocity channel with a beamsize of 30$''$. WKK~6801 and WKK~6489
represent thus 9$\sigma$ and 3$\sigma$ detections.
If we assume an average M$_{\rm HI}$/L$_{\rm B}$=0.2~M$_{\odot}$/L$_{\odot,{\rm B}}$
(Roberts \& Haynes 1994) and a limiting H{\sc i} mass of 7\,10$^{8}$~M$_{\odot}$
(see Sect.~\ref{sec:detect}), we obtain a limiting luminosity
L$_{B} \simeq 3.5\,10^{9}$~L$_{\odot, {\rm B}}$. With a distance of 79~Mpc this
leads to B$_{\rm J}^{0}$=16.2. The number of cluster spiral galaxies with
B$_{\rm J}^{0} \leq$16.2 within the Abell radius is 214 (Woudt 1998).
The fraction between the search area (the three inner fields) and
the area within an Abell radius is $\Delta \sim 0.05$. If the spiral galaxies in
ACO~3627 had M$_{\rm HI}$/L$_{\rm B}$ comparable to those of field galaxies,
we should have detected more than 10 galaxies.

\section{Discussion}

\subsection{Detection rate \label{sec:detect}}

With an H{\sc i} column density sensitivity limit of
$\sim 2\,10^{20}$ cm$^{-2}$ we are not able to detect
dwarf galaxies nor low surface brightness galaxies (Pickering et al. 1997)
in our fields. The galaxies we expect to be detectable will thus have
linewidths of at least $W \sim$100 km\,s$^{-1}$.
We can estimate the H{\sc i} mass detection
limit in the following way: an assumed flat-topped spectrum
of intrinsic width $W$, due to the galaxy's inclination
has an observed velocity range of
$\tilde{W} \simeq W\big(1-(\frac{d}{D})^{2}\big)^{\frac{1}{2}}$,
where $D$ is the optical diameter along the major axis and $d$ is the
minor axis diameter.  We further assume a spatial extent of one beamsize
or less per velocity channel.
At a distance of 79 Mpc this implies an H{\sc i} mass limit of
\begin{equation}
M_{HI} \simeq \tilde{W} \times 3\,10^{-3}  \times
2.36\,10^{5}\times 79^{2}\, {\rm M}_{\odot}\ ,
\end{equation}
where $\tilde{W}$ is in km\,s$^{-1}$.
Assuming $\tilde{W}$=150 km\,s$^{-1}$ gives a limit of
$\sim$7\,10$^{8}$ M$_{\odot}$, for example.

We can estimate the H{\sc i} linewidth of individual objects using the B-band
Tully Fisher (TF) relation. Adopting the Virgo cluster B-band TF relation
found by Yasuda et al. (1997), and using
a distance of 17 Mpc for the Virgo cluster and 79 Mpc for ACO~3627,
we obtain as TF relation for ACO~3627:
\begin{equation}
B_{\rm T}^{0}=-7.92({\rm log}\,W_{20}-2.5)+15.62
\end{equation}
We have calculated a limiting H{\sc i} mass-to-blue light ratio,
M$_{\rm HI}$/L$_{\rm B}$, for each individual spiral galaxy identified
in our three ACO~3627 ATCA fields by Woudt (1998),
using the linewidth obtained with the TF relation and relation (1).

The high extinction makes it difficult to identify the exact morphological
type of any individual object. Woudt (1998) made an attempt to
classify the detected galaxies where it was possible. We decided to separate
the galaxies detected and classified by Woudt (1998) into three different
groups according to their morphology:
group I: 22 early-type galaxies (S0/a--Sb),
group II: 15 late-type galaxies (Sbc--Sd), and
group III: 31 spiral galaxies without further morphological subclassification (S type).
The mean limits to the mass-to-light ratio and their
dipersions for these groups are
\begin{itemize}
\item
Group I:\,\, M$_{\rm HI}$/L$_{\rm B}$=0.17$\pm$0.10 M$_{\odot}$/L$_{\odot, \rm B}$\item 
Group II:\,  M$_{\rm HI}$/L$_{\rm B}$=0.22$\pm$0.09 M$_{\odot}$/L$_{\odot, \rm B}$\item 
Group III:   M$_{\rm HI}$/L$_{\rm B}$=0.34$\pm$0.11 M$_{\odot}$/L$_{\odot, \rm B}$      
\end{itemize}
The global limit to the mass-to-light ratio for all galaxies is
M$_{\rm HI}$/L$_{\rm B}$=0.25$\pm$0.13 M$_{\odot}$/L$_{\odot, \rm B}$.

For comparison, Roberts \& Haynes (1994) give the following average
H{\sc i} mass-to-blue light ratios and dispersions as function of
morphological types:
S0:    0.03$\pm$0.02 M$_{\odot}$/L$_{\odot, \rm B}$,
Sa/Sb: 0.1$\pm$0.07  M$_{\odot}$/L$_{\odot, \rm B}$, and
Sc/Sd: 0.3$\pm$0.15  M$_{\odot}$/L$_{\odot, \rm B}$.
Our mean limit to the mass-to-light ratio for the late-type (Sbc--Sd)
galaxies in ACO~3627 is below the mean value given for Sc-Sd spirals
by Roberts \& Haynes.
Moreover, we find 8 galaxies (or 36\%) in group I (S0/a--Sb) with
M$_{\rm HI}$/L$_{\rm B} \leq 0.1$ M$_{\odot}$/L$_{\odot, \rm B}$,
the average for Sa-Sb spirals found by Roberts \& Haynes,
and 7 galaxies (or 47\%) in group II (Sbc--Sd) with
M$_{\rm HI}$/L$_{\rm B} \leq 0.2$ M$_{\odot}$/L$_{\odot, \rm B}$,
while 0.3 is the mean value for Sc-Sd spirals by Roberts \& Haynes.
We should have detected at least some of the spirals identified
in the ACO~3627 cluster, if they have M$_{\rm HI}$/L$_{\rm B}$ ratios
comparable to those found by Roberts \& Haynes.

In principle there are three possible explanations why they are not detected:
\begin{enumerate}
\item
The galaxies with unknown velocities are outside the velocity range.
This is, however, quite unlikely considering the velocity range
covered with the ATCA observations being ($2300 - 7700$ km\,s$^{-1}$)
compared to the mean velocity and velocity dispersion of ACO~3627
of $<{\rm V}>$ = 4844 km\,s$^{-1}$, respectively $\sigma_{\rm V}$ =
848 km\,s$^{-1}$ (cf. also Column 9 in Table~\ref{tab:galaxies}).
\item
Some early type galaxies are misclassified as spirals.
Since it is very difficult to determine morphological types
due to the high extinction in the Galactic plane, misclassifications
are always possible. Nevertheless, there should not be a systematic
trend to misclassify lenticular galaxies with intrinsically featureless
disks as spirals.
\item
They are H{\sc i} deficient.
\end{enumerate}
We conclude that even if we take the two former points into
account there is evidence that a significant fraction of the
central spiral galaxies are H{\sc i} deficient.

\subsection{Comparison with the Coma cluster detection rate\label{sec:coma}}

The X-ray luminosity, velocity dispersion, galaxy magnitude
distribution of ACO~3627 all are comparable to that of the Coma cluster (Woudt
1998).  We will therefore compare our detection rate with that
measured with the VLA
in Coma by Bravo-Alfaro et al. (2000). Even if their data do not cover
the whole velocity range of the cluster, they should not have missed
a considerable number of galaxies, because their fields where centered
on the galaxies with the highest B luminosity (B$_{\rm T}^{0}<15.7$).
They detected 19 galaxies in a search
area of $\sim$2.3 $\Box^{\rm o}$ with an rms noise of 0.4 mJy/beam.
With our sensitivity ($\sim$2\,10$^{20}$ cm$^{-2}$) they would have
detected 9 galaxies.
This gives a detection rate of $n_{\rm Coma}$=3.9 galaxies/$\Box^{\rm o}$.
We have detected 2 galaxies in a search area of $a=$0.75 $\Box^{\rm o}$
(the three central fields) leading to  $n_{\rm A3627}$=2.67
galaxies/$\Box^{\rm o}$.

In the case of spatially resolved H{\sc i}
emission the H{\sc i} flux density of galaxies of the same physical size and
the same total H{\sc i} mass is the same for both clusters.
Therefore, we only have to correct
the Coma detection rate for the search area in ACO~3627.
If the H{\sc i} emission is not resolved an additional
correction for beam dilution effect must be taken into account.
Since the galaxies in both clusters should be spatially
resolved in H{\sc i}, we will consider only the first case.
If we correct the Coma detection rate for the distance of ACO~3627 we get
\begin{equation}
n^{\rm corr}_{\rm Coma}=n_{\rm Coma}\,
(\frac{\langle v \rangle_{\rm A3627}}{\langle v \rangle_{\rm Coma}})^{2}=2.0
\ {\rm galaxies/}\Box^{\rm o},
\end{equation}
where $\langle v \rangle_{\rm A3627}$=4844
km\,s$^{-1}$ and $\langle v \rangle_{\rm Coma}$=6853 km\,s$^{-1}$.
The expected number of detection for our survey is
$n^{\rm ex}_{\rm A3627}=n^{\rm corr}_{\rm Coma}\times a \simeq 1.5$ galaxies.
Thus, we have a reasonable detection rate compared to that of the Coma
cluster.

Moreover, out of 21 galaxies in the three inner fields with
B$_{\rm T}^{0}<15.2$ (this
corresponds to the B$_{\rm T}^{0}<15.7$ in the Coma cluster) we detected two
galaxies. Bravo-Alfaro et al. (2000) would have detected 9 galaxies out
of 44 in the Coma cluster with our sensitivity. On the other hand,
the cluster ACO 262 located at nearly the same distance
as ACO 3627 and similar to the Virgo cluster, i.e. only mildly H{\sc i}
deficient, has been mapped with the Westerbork array by Bravo-Alfaro et al.
(1997). They detected 11 out of 25 galaxies
(B$_{\rm T}^{0}<15.2$) with a sensitivity comparable to ours.
Clearly, the H{\sc i} detection rate in ACO~3627 is similar to that of
the H{\sc i} deficient Coma cluster and not that of A~262, indicating
that the bright spiral galaxies in ACO~3627 are H{\sc i} deficient.

\section{Conclusions}

We have observed three different fields of 30$'$ diameter each
within the Abell radius of the galaxy cluster ACO~3627 as well as
one comparison field outside the cluster  with an rms noise of
$\sim$1 mJy/beam. We have detected 2 galaxies for the first time:
WKK~6801, located at a projected distance of 2.6$^{\rm o}$=3.5 Mpc
east of the cluster center, seems to be a very late type spiral.
For WKK~6489, located 1$^{\rm o}$=1.3 Mpc south east of the cluster center,
our data does not permit to draw firm conclusions about its exact nature
but it is probably an H{\sc i} deficient galaxy.

The detection rate is in reasonable agreement with that
obtained for the Coma cluster (Bravo-Alfaro et al. 2000).
There is evidence that a large fraction of spiral galaxies
in the inner $1^{\rm o} \times 1^{\rm o}$ area are H{\sc i} deficient.
This deficiency is probably due to the interaction of the atomic
gas in the galaxies with the hot intracluster gas detected in
X-rays by B\"{o}hringer et al. (1996). We therefore expect
that ram pressure stripping is a probable cause for the
H{\sc i} deficiency of the spiral galaxies in ACO~3627.

\begin{acknowledgements}
We particularly would like to thank B. Koribalski for doing one observing run
and for giving important advise on the observing procedure.
We also wish to thank the ATCA and ATNF staff for their kind support during the
observations.\\
BV is supported by a TMR Programme of the European Community
(Marie Curie Research Training Grant). The research of P.A.H. is supported
by NSF Faculty Early Career Development (CAREER) Program award AST 95-02268.
\end{acknowledgements}

\end{document}